\documentclass{PoS}
\usepackage{amsmath,amssymb}
\usepackage{tikz-feynman}
\tikzfeynmanset{compat=1.1.0}

\usepackage{tikzsymbols}

\title{Leptogenesis from Dark Matter Annihilations in Scotogenic Model}

\ShortTitle{Leptogenesis in minimal Scotogenic model}

\author{Debasish Borah\\
        Department of Physics, Indian Institute of Technology, Guwahati, Assam 781039, India\\
        E-mail: \email{dborah@iitg.ac.in}}

\author{\speaker{Arnab Dasgupta}\\
        School of Liberal Arts, Seoul-Tech, Seoul 139-743, Korea\\
        E-mail: \email{arnabdasgupta@protonmail.ch}}

\author{Sin Kyu Kang\\
        School of Liberal Arts, Seoul-Tech, Seoul 139-743, Korea\\
        E-mail: \email{skkang@seoultech.ac.kr}}

\abstract{We address the possibility of realising successful leptogenesis from dark matter annihilations in the scotogenic model of neutrino masses to explain the same order of magnitude abundance of dark matter and baryons in the present Universe.
After showing that the minimal model in this category can not satisfy all these requirements, we study a minimal extension of this model and find that the scale of leptogenesis can be as low as 5 TeV, lower than the one in vanilla leptogenesis scenario in scotogenic model along with the additional advantage of explaining the baryon-dark matter coincidence. Due to such low scale, the model remains predictive at dark matter direct detection and rare decay experiments looking for charged lepton flavour violating processes.}

\FullConference{ICHEP 2018, International Conference on High Energy Physics\\
5-12 July 2018, Seoul, South Korea}

\begin{document}

\section{Failure of Vanilla Scotogenic Model}
The scalar DM annihilations can produce a net lepton asymmetry after the annihilation rates go out of equilibrium.

\begin{figure}
\centering
\begin{tabular}{lr}
\begin{tikzpicture}[/tikzfeynman/small]
\begin{feynman}
\vertex (i){$\eta^{0,-}$};
\vertex [below = 1.6cm of i] (j){$\eta^{0,-}$};
\vertex [right = 1.6cm of i] (k);
\vertex [right = 1.6cm of k] (l){$\nu,l^-$};
\vertex [right = 1.6cm of j] (m);
\vertex [right = 1.6cm of m] (n){$\nu,l^-$};
\diagram*[small]{(i) -- [charged scalar] (k),(k) -- [fermion] (l),(j) -- [charged scalar] (m),(m) -- [fermion] (n),(k) -- [majorana,edge label =$N$] (m)};
\end{feynman}
\end{tikzpicture}

\begin{tikzpicture}[/tikzfeynman/small]
\begin{feynman}
\vertex (i){$\eta^{0,-}$};
\vertex [below = 1.6cm of i] (j){$\eta^{0,-}$};
\vertex [right = 0.8cm of i] (v1);
\vertex [below = 0.8cm of v1] (v);
\vertex [right = 0.8cm of v1] (k);
\vertex [right = 1.6cm of k] (l){$\nu,l^-$};
\vertex [right = 0.8cm of j] (v2);
\vertex [right = 0.8cm of v2] (m);
\vertex [right = 1.6cm of m] (n){$\nu,l^-$};
\diagram*[small]{(i)--[charged scalar] (v),(v) -- [charged scalar] (k),(k)--[fermion] (l),(j)--[charged scalar] (v),(v) -- [charged scalar] (m),(m)--[fermion] (n),(k) -- [majorana,edge label =$N$] (m)};
\end{feynman}
\end{tikzpicture}

\end{tabular}
\caption{Feynman diagrams corresponding to the annihilation/co-annihilation processes in $\langle\sigma v\rangle_{\rm CDM\; CDM \rightarrow L L}, \langle\sigma v\rangle_{\rm CDM \;CDM \rightarrow L \;WDM}$.}
\label{fig1}
\end{figure}
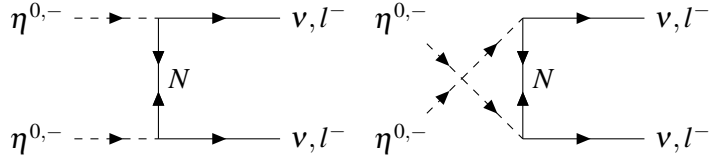
Now the major downfall of the model comes from the Direct Detection constraints arising from tree level $Z$ boson mediated processes $\eta_R n \rightarrow \eta_I n$, $n$ being a nucleon. Which gives a bound of about
$
\lambda_5 \approx 1.65 \times 10^{-7} \left( \frac{\delta}{100 \; \rm keV} \right) \left( \frac{M_{\rm DM}}{100 \; \rm GeV} \right). \nonumber
$
Now in order to satisfy the above relation we would require lower $\lambda_5$ which in turn would conflict with lower bound required for the neutrino mass. In order to alleviate this tension we need to incorporate another singlet scalar.

\section{Conclusion}

We have addressed the possibility of explaining the coincidence of DM abundance and baryon asymmetry in the present Universe along with non-zero neutrino masses
within the framework of scotogenic model.  Adopting the WIMPy leptogenesis framework and considering the minimal scotogenic model, we first show that the simultaneous generation of DM abundance and baryon asymmetry along with satisfying neutrino mass, DM direct detection data is not possible.  We then minimally extend the model by another complex singlet\cite{Arhrib:2015dez} and found that it is indeed possible to explain $\Omega_{\rm DM} \approx 5 \Omega_{B}$ along with satisfying all other phenomenological constraints.  For the two benchmark points chosen in our work, we could obtain successful leptogenesis along with other requirements for $M_{\rm DM} \sim 5$ TeV whereas vanilla leptogenesis in scotogenic model works for $M_R \geq 10$ TeV\cite{Hugle:2018qbw}.  The model can still be tested at near future run of direct detection experiments like Xenon and rare decay experiments looking for charged lepton flavour violation like $\mu \rightarrow e \gamma, \mu \rightarrow 3 e$, $\mu$ to $e$ conversion etc.


\begin{thebibliography}{99}

\bibitem{Ma:2006km}
E.~Ma, {\it {Verifiable radiative seesaw mechanism of neutrino mass and dark
  matter}},  {\em Phys. Rev.} {\bf D73} (2006) 077301,
  [\href{http://arxiv.org/abs/hep-ph/0601225}{{\tt hep-ph/0601225}}].

\bibitem{Arhrib:2015dez}
A.~Arhrib, C.~Boehm, E.~Ma, and T.-C. Yuan, {\it {Radiative Model of Neutrino
  Mass with Neutrino Interacting MeV Dark Matter}},  {\em JCAP} {\bf 1604}
  (2016), no.~04 049, [\href{http://arxiv.org/abs/1512.08796}{{\tt
  arXiv:1512.08796}}].

\bibitem{Borah:2018uci} 
  D.~Borah, A.~Dasgupta and S.~K.~Kang,
  arXiv:1806.04689 [hep-ph].

\bibitem{Hugle:2018qbw} 
  T.~Hugle, M.~Platscher and K.~Schmitz,
  Phys.\ Rev.\ D {\bf 98}, no. 2, 023020 (2018)
  doi:10.1103/PhysRevD.98.023020
  [arXiv:1804.09660 [hep-ph]].
\end{thebibliography}
\end{document}